A New Approach to Reporting Archaeological Surveys: Connecting Rough Cilicia,

Visible Past and Open Context through loose coupling and 3d codes

Sorin Adam Matei*
Brian Lamb School of Communication
BRNG 2132
Purdue University
100 N. University Drive
West Lafayette, IN 47907

Eric Kansa
Alexandria Institute
125 El Verano Way
San Francisco, CA 94127

Nicholas Rauh
School of Languages and Cultures
Stanley Coulter Hall, Purdue University
W. Lafayette, IN 47907

*corresponding author

Abstract

Background: The project presents the strategy adopted by the Rough Cilicia Archaeological Survey team for publishing its primary data and reports via three potentially transformative strategies for digital humanities: 1. Loose coupling of digital data curation and publishing platforms. In loosely coupled systems, components share only a limited set of simple assumptions, which enables systems to evolve dynamically. 2. Collaborative creation of map based narrative content. 3. Connecting print scholarship (book, reports, article) to online resources via two-dimensional barcodes (2D codes) that can be printed on paper and can call up hyperlinks when scanned with a Smartphone.

Case study: The present essay provides a higher level framework for developing digital humanities in in general, via three possible strategies generated by loosely coupling two autonomous services: Visible Past, dedicated to web collaboration and digital-print publishing and Open Context, which is a geo-historical data archiving and publishing service.

Conclusions: The Rough Cilicia Archaeological Survey, Visible Past, and Open Context work together to illustrate a new genre of scholarship, which combine qualitative narratives and quantitative representations of space and social phenomena. The project provides tools for collaborative creation of rich scholarly narratives that are spatially located and for connecting print publications to the digital realm. The project is a case study for utilizing the three new strategies for creating and publishing spatial humanities scholarship more broadly for ancient historians.

Introduction

The free and open circulation of ideas lies at the core of humanistic scholarship. The emergence of digital and online tools for supporting humanities scholarship can considerably speed up this ideal. Yet, as the number of publishing, mapping, and data management platforms proliferate, many technical and methodological incompatibilities may also appear, endangering the ideal of open communication. Ultimately, scholars need practical methods to connect them so that information can efficiently flow across platforms to generate new scholarship. Removing potential walls between narratives, primary data, and maps holds particular importance for the spatial humanities. Narratives should naturally follow the contours of landscapes and speak to the reader through the power of maps. At the same time, as the publishing world shifts from static, paper-based, to online digital platforms, established practices for evaluating and disseminating humanities endeavors should be continued, even as new publication methods are encouraged.

The present project proposes a possible solution for these challenges by using a "loose coupling" strategy for supporting the Rough Cilicia Archaeological Survey project [1,2]. The project, which is at the framework stage, does not aim to test the superiority of the proposed method for enhancing collaboration, but to delineate principled methods by which collaboration can be supported. The specific workflow by which collaboration can be enhanced is still to be developed and appropriate usability testing should sand will be employed to ascertain its success.

Background

The purpose of the Rough Cilicia Archaeological Survey Project is to examine the process of Roman provincial acculturation through the lens of Rough Cilician material and cultural remains. Conceived as a regional survey in western Rough Cilicia (a mountainous coastal region along the south coast of modern Turkey, modern Antalya Province, Gazipasha District), the Rough Cilicia team has investigated some 300 km2 of archaeological terrain in thirteen seasons (1996-2011). The team has mapped and processed surface remains at approximately 150 "loci" (including 10 small urban sites) of past human activity. It has constructed plans for 12 selected built environments ranging from large urban settlements along the coast (Selinus, Iotape, Kestros, Antioch) to isolated farms in the hinterland. It has compiled a dataset of 8350 artifacts (mostly ceramic remains) individually georeferenced and hot-linked to digital photographs, tens of hours of video recording, and hundreds of pages of field notes. Completed in 2011, the project now faces the daunting task of organizing its array of data in an accessible manner. We are attempting to utilize two online platforms, Visible Past and Open Context (Matei, Kansa, Rauh, 2011) to store and publish archaeological data gathered by the survey Project. Before presenting an example of the manner in which Rough Cilicia is being served by the two platforms, a few words about each of them:

Visible Past: Visible Past [3,4] (http://visiblepast.net) is a publication platform that brings together narratives, primary data, and dynamic Web maps. Content can be viewed online, but also accessed in other formats, including electronic and print-on-demand books. The latter are enhanced with 2d codes, which link paper and online resources [5] (see http://matei.org/url/virtsoc. Click "Look inside" to view book pages).

Open Context: Open Context [6,7,8] (http://opencontext.org) publishes editorially reviewed primary data for archaeology and related fields. Special workflows improve the quality of data contributed by researchers by using appropriate technical and semantic standards. Open Context's Web-services transmit these data to other services, such as Visible Past, via simple data feeds in widely supported standards (Atom, KML, JSON, etc). All data is archived with the California Digital Library and citable with persistent identifiers. Creative Commons licenses enable anyone to analyze, repurpose and visualize Open Context data. Both the National Science Foundation's (NSF) Archaeology Program and the National Endowment for the Humanities (NEH) reference Open Context for newly required "Data Management Plans".

Visible Past and Open Context work together to meet the challenge of "spatial humanities" [9], which strive to combine qualitative evidence and quantitative representations of space and social phenomena. Within spatial humanities environments narratives and qualitative evidence remain the driving forces, while spatial representation and data are their scaffolds. Visible Past and Open Context attempt to meet the challenge of digitizing and mapping humanistic knowledge through three strategies: loose coupling, deep mapping, and trans-media publishing [10]. The main publishing platform, Visible Past is designed to mesh maps and narratives—articles, book chapters, or reports—in a natural way through its three focal points: map, narrative, and interactive/comment area. Each unit of content is shown on a map together with any primary data and/or 3D reconstructions. Open Context's archives primary data and makes it available through data streams (Atom, KML, JSON) that are easy to query, capture and reuse across the Web. Widely used Web standards and design patterns, particularly "RESTful" (Web-style) service design [8] enables data from Open Context to dynamically flow into other Web-based systems. We call this approach "loosely coupled," because various component systems can evolve independently and can be swapped in or out for other systems with greater ease. In our example, Visible Past presents feeds of data from Open Context, specifically filtered and selected for relevance to the narrative argument presented in Visible Past. In principle however, Visible Past can similarly accept data streams from other standards-compliant sources, and Open Context can similarly present data for use in platforms other than Visible Past.

Case study

To illustrate the points made above we will present a case study of implementing the loose coupling framework for developing digital humanities platforms. The case study is in this case the "method" by which we hope to make the case for our proposition. In what follows we will present the workflow by which Rough Cilicia data has been brought from the field to the end user via Visible Past and Open Context. We use as an example a sub-sample of recorded architectural features, tombs or funerary monuments that have been investigated by the Rough Cilicia Archaeological Survey Project. Over 13 field seasons, the survey has assembled records for 231 funerary monuments or tombs at 26 discrete and named locations. These include a wide variety of tombs, including temple tombs, grabhaus or mausoleum tombs, mortar constructed, vaulted chambers, rock cut tombs, smaller Lycian-style house tombs, pedestals, altars, and stone carved ossuary boxes knows as larnaces or ostothekai. Over the course of time the team noticed that the locations of the various types of tombs varied with respect to their proximity to the sea or the Tauros Mts (some 30 km. to the interior), and that this variation of arrangement was equally determinable by altitude, as well as by the type of site (urban, village, isolated necropolis). Put simply, tomb designs that were not native to the region, that is, tomb types that were not present before the Roman era (end first century BC) and borrowed heavily from mainstream, off-shore, Greco-Roman building design and technology tended to be situated closer to the sea. These include monumental structures such as temple tombs, grabhaus, and mortar-built chamber tombs, since all of these depended on design and construction techniques that were incorporated into regional tomb construction during the Early Roman era. Rock-cut tombs with dressed reliefs to mimic house facades, human busts in intaglio (often with inscribed memorials) and Lycian house tombs appear to have been more commonplace in the Mesogeia or Midlands of south coastal Anatolia. As opposed to the rigidly family-oriented tomb memorials along the coast, many of the rock cut tombs at midland sites such as Lamos and Direvli record the jointly contributed tomb development of unrelated inhabitants, something that we have referred to as corporate tombs.

Similar tombs are visible in Lycia (naturally) and eastern Rough Cilicia, though there are discernible variations. The smallest worked tomb structures, larnaces, pedestals, and altars used as tomb markers tend to be arranged in necropolis clusters found only at the Midland sites, usually in a low place situated at the entrance to a cliff-top settlement such as Kenetepe (RC 0304) or Corus (RC 1105). The closest comparison for these necropolis clusters are found in the Isaurian hinterland of the Tauros Mts.; hence their placement here seems to reflect the extended reach of Isaurian cultural attributes toward the Mediterranean coast. Since the necropolis clusters seem to be found only at Midland sites they seem to indicate that the Isaurian influence did not extend to the coast. Despite overwhelming evidence that the coastal inhabitants were as autochthonous as those of the hinterland, these inhabitants appear rather to have assimilated mainstream Greco-Roman cultural attributes in their tomb construction.

The team decided to organize these variables into a as dataset to see whether spatial analysis would confirm the presence of such a pattern of distribution. Data was classified not only by type, but also by the urban vs. rural context or by degree of isolation (necropolis vs. isolated). Analysis of pattern of distribution for funerary sites and features indicates three clear cultural areas. Figure 1 displays an attempt to interpolate cultural areas from the spatial distribution of the funerary sites uncovered by the Rough Cilicia Archaeological Survey.

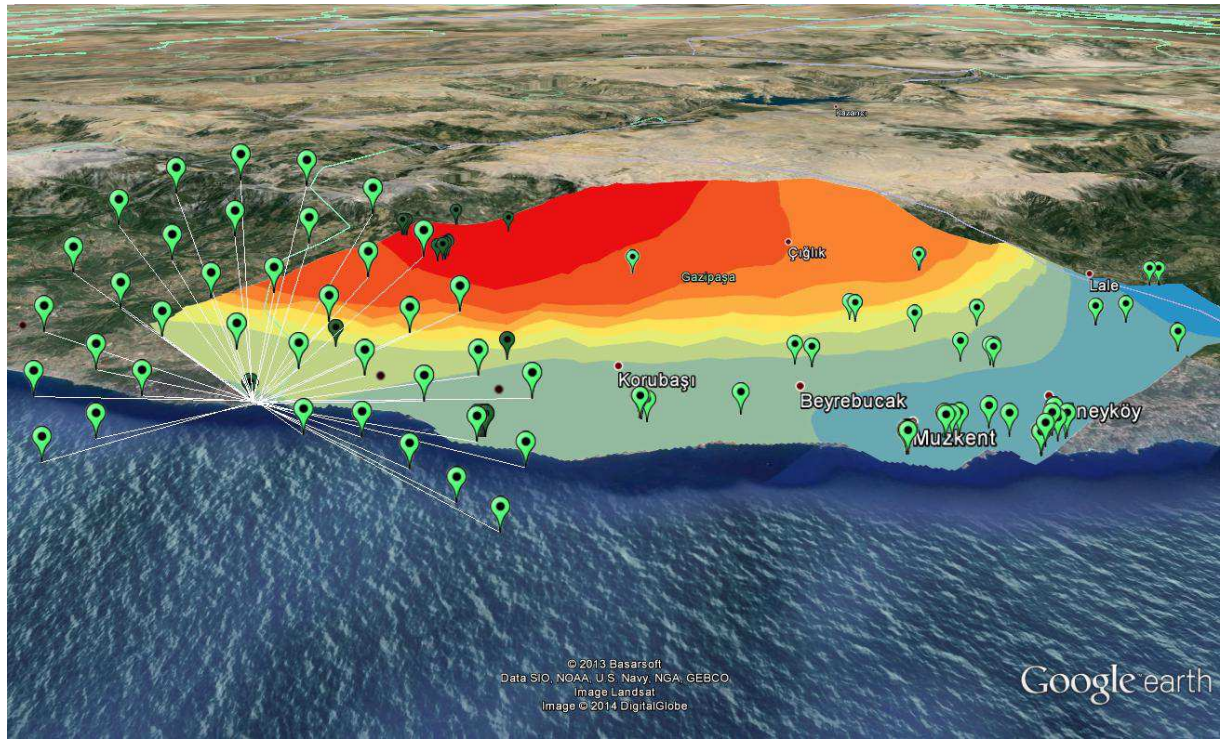

Figure 1. Cultural zones in Rough Cilicia interpolated from funerary features. Map data: Google.

Bands of color represent interpolated values for probable cultural areas. Red bands indicate higher likelihood of hinterland cultural affiliation. Blue bands indicate higher likelihood of coastal/polis cultural affiliation. Orange/yellow bands indicate higher likelihood of Mesogeia cultural affiliation. The map in Figure 1 clearly indicates that cultural types change with distance from coast and altitude and that discrete areas of cultural stratification can be discerned spatially. The procedure used was diffusion interpolation with barriers. Analysis starts with assigning each feature to a cultural type: polis – highly acculturated, mesogeia – medium acculturated, and hinterland – weakly acculturated. Classification criteria included technological, architectural, decorative, and symbolic characteristics. Once divided into acculturation types, diffusion interpolation determines areas of equal probability where features assigned specific cultural types could be found. To make sure that features found at the edges of the study area do not have an undue influence on the analysis, we used as a barrier a conventional search border that delineates the survey sector involved in this study. The color bands represent interpolated values for probable cultural areas. The red bands indicates higher likelihood of hinterland cultural affiliation (low acculturation). As can be noticed, it appears in the form of a mountain cap, following the contours of higher elevations of the coastal mountains and of the hinterland plateau. The orange and yellow bands indicate higher likelihood of Mesogeia cultural affiliation (semi-acculturation). It generally keeps a constant 7 km distance from the coast, except for the north-west portion, where it meets the sea. This is due to the fact that elevations in that area remain elevated even in the immediate vicinity of the coast, denying high acculturation, polis-like habitations the space and locations they favor. The blue and green bands indicate higher likelihood of coastal/polis cultural affiliation, which follow the contour of the coast, except for the lower south-east corner, where they bend inland, following the Kaledran Cayi valley. As is expected, riveran basins tend to be more hospitable to high acculturation locales. Furthermore, the bands are not simply stating the obvious. As could be seen on the map, each of the three bands may include funerary features that are not "specific" to it. These, are, however, exceptions, and statistical analysis shows that their presence in that area is determined by local factors that slightly diverge from the general trend, as is expected with all human related behaviors.

The next step is to present this data in an intelligible, accessible manner. The data was mounted on Open Context, where the variables (urbanization, isolation, construction technology, typology, inscriptions, photographs, etc.) became as many searchable facets. Once the records are mounted in Open Context, the funerary and tomb artifact database becomes more than a static or citable record-holding archive. It becomes a dynamic information service. Records can be searched, sorted, and reutilized in discrete batches, each pertaining to a specific documentation need. Open Context also aligns data with common semantic concepts. In this case, the Rough Cilicia results will be related to the Pleiades Gazetteer (http://pleiades.stoa.org/), managed by the Institute for the Study of the Ancient World at New York University. Using "Linked Open Data" methods (using Web identifiers to reference shared concepts, in this case geographic "place"), Rough Cilicia results can be further contextualized by a growing body of open data published by many museums and digital collections now referencing Pleiades (see http://pelagios-project.blogspot.com/). Open Context's participation in the growing information ecosystem of Linked Open Data makes it easier to precisely define relevant streams of data from Open Context that can be used in other Web applications, including Visible Past.

Once data is stored in Open Context and hyperlinked to other "Linked Open Data" sources, researchers can search it using the predefined variables. A novel and powerful feature offered by Open Context is the capability to integrate the search interface and the results, which are served as either lists or maps, on third party sites, such as Visible Past. The mock up report at http://visiblepast.net/see/archives/1380 summarizing the 231 funerary features discovered by the Rough Cilicia project illustrates this concept. Upon clicking the markers in the main map, records are automatically downloaded from Open Context and presented on Visible Past. Visible Past can freely interconnect with Open Context through open Web services. Instead of being a simple collection of html files, Visible Past relies on WordPress, the popular online software that stores the content in a database. When Visible Past authors create new pages (reports, papers, maps, etc.) they do so via a word processor-like interface. However, the content is stored as a complex database. Each page is created when the viewer requests it and is enhanced with local and remote information (maps, pictures, videos, etc.). As new resources are added to the site, the pages are automatically updated. The tool by which authors can search and utilize on Visible Past data and maps from Open Context is embedded as an "instrument panel" in the edit interface of each article allowing searches by site, type of artifact, or very precise geographic location.

The figure above offers a snapshot of this tool as is seen on Visible Past. The results of the selection process are passed from Open Context to Visible Past, where they become available as a .KML file, an open source file format for creating interactive maps. By default, the datasets selected from Open Context will show in a master map, which is associated with each Visible Past article. However, spatial data can also be linked and called from mini-maps, embedded in the body of the narrative. When clicked, these mini-maps turn into full screen maps (See above). Data can be examined both in two and three dimensions (on Visible Past click Earth to see the maps in 3D). Furthermore, each record displayed on the map is back-linked to Open Context, which allows the readers of the narrative to examine the records in more detail in the original data-publishing context.

Once finalized, the narratives can be repackaged as ebooks, downloadable in the popular epub Ipad or Kindle formats. The maps can be hyperlinked even in the ebook format and when clicked, will bring up any mapping interface associated to the two platforms. Furthermore, the narratives can be repackaged as pdf files, which can then be used on most desktops or mobile devices or can be passed on to a print on demand service, such as Amazon.com, to be delivered as a traditional publication, available to libraries or bookstores throughout the world. Pdf files and print publications will include 2d codes, such as the one to the left, which will provide hyperlink access to Visible Past even when reports are printed and not available in electronic format  (You may download a demonstration example of the report in a print-ready format at http://visiblepast.net/see/wp-content/uploads/2012/04/RoughCiliciaExample2dCodeEnhanced.pdf ). The scanning of 2d codes and downloading of information is intermediated by mobile devices. A practical demonstration of these features will be offered at the conference. An example of a 2d code enhanced book produced by similar methods by one of the co-authors of this paper can be found at http://matei.org/url/17s .

Conclusions

Our vision is to use Visible Past and Open Context to connect the interactive maps, field notes, survey narratives, videos, and primary data of the Rough Cilicia Archaeological Survey Project (such as the spatial tomb data presented here) and to publish this online as well as a 2d code enhanced book. The loose coupling of Visible Past/Open Context, the deep mapping, and print-digital publishing tools they employ offer Rough Cilicia archaeologists the means needed to work together virtually and to integrate publishing or archiving of spatial data and narratives. Our loosely coupled platform also supports collaboration by minimizing bottlenecks created by standards incompatibilities. At the same time, it makes data dissemination and preservation more resilient by promoting saving and disseminating of information from multiple repositories and publishing centers. Although this seems redundant and inefficient, a simple cost benefits analysis should reveal to be a net benefit for digital humanities projects. Many projects have moderate to short life spans. Unless supported by major funders, they tend to fade away and be lost. Storing the data separate from the narratives, especially via well funded, resilient platforms, such as Open Contexts which uses for archiving the California Digital Library, will ensure visibility and longer life spans for primary data.  Rather than building monolithic silos, a loosely coupled approach also enables us to leverage an ecosystem of software and information services on the Web to meet different publishing and visualization needs.  Since the needs of our project are not unique to field-based archaeology, our project holds wider multidisciplinary relevance. Our experiments will inform the development of similar spatially-oriented projects by helping to identify key interface, semantic, and data quality requirements and good practices.

SAM, Associate Professor of Communication, Director of Research for Computational Social Science, Cyber Center, Purdue University

NHK, Professor of Classics, Purdue University

ECK, Alexandria Archive founder, Open Context Editor